\documentclass[aps,twocolumn,superscriptaddress,amsmath,amssymb,floatfix]{revtex4-1}

\usepackage{graphicx,tabularx}
\usepackage{dcolumn}
\usepackage{bm}
\usepackage{color}

\usepackage{upgreek}

\usepackage{epstopdf}

\begin{document}


\title{Nanometers-thick Ferromagnetic Surface Produced by  Laser Cutting of  Diamond}

\author{Annette Setzer}
\author{Pablo D. Esquinazi}\email{esquin@physik.uni-leipzig.de}
\affiliation{Division of Superconductivity and Magnetism, Felix-Bloch-Institute for Solid State
Physics, University of Leipzig, 04103 Leipzig, Germany}

\author{Sergei Buga}\altaffiliation{Also at Moscow Institute of Physics and Technology, 9
Institutskiy per., Dolgoprudny, Moscow Region, 141701 Russia}
\affiliation{Technological Institute for Superhard and Novel Carbon Materials,
 7a Centralnaya street, Troitsk, Moscow, 108840 Russia}

\author{Milena T. Georgieva}\altaffiliation{On live from: Condensed Matter Physics and Microelectronics Division, 
Faculty of Physics, Sofia University St. Kl. Ohridski, 
James Bouchier Blvd 5, 1164 Sofia, Bulgaria }
\affiliation{Division of Superconductivity and Magnetism, Felix-Bloch-Institute for Solid State
Physics, University of Leipzig, 04103 Leipzig, Germany}

\author{Tilo Reinert}
\affiliation{Division of Applied Quantum Systems, Felix-Bloch-Institute for Solid State
Physics, University of Leipzig, 04103 Leipzig, Germany}

\author{Tom Venus}
\author{Irina Estrela-Lopis}
\affiliation{Institute of Medical Physics and Biophysics, University of Leipzig, D-04107 Leipzig, Germany}

\author{Andrei Ivashenko}
\author{Maria Bondarenko}
\affiliation{Technological Institute for Superhard and Novel Carbon Materials,
 7a Centralnaya street, Troitsk, Moscow, 108840 Russia}

 \author{Winfried B\"ohlman}
 \affiliation{Division of Superconductivity and Magnetism, Felix-Bloch-Institute for Solid State
Physics, University of Leipzig, 04103 Leipzig, Germany}

 \author{Jan Meijer}
 \affiliation{Division of Applied Quantum Systems, Felix-Bloch-Institute for Solid State
Physics, University of Leipzig, 04103 Leipzig, Germany}

black
\begin{abstract}
 In this work, we demonstrate that cutting diamond  crystals with a laser (532~nm wavelength, 0.5~mJ energy, 200~ns pulse duration at 15~kHz) 
produces a $\lesssim 20~$nm thick surface layer with
 magnetic order at room temperature. We have measured the magnetic moment   
  of five natural   and six CVD diamond crystals 
of different size, nitrogen content and surface orientations with a SQUID magnetometer. 
A robust ferromagnetic response at 300~K is observed  only for crystals that were cut with  the laser  along the (100) surface orientation. 
The magnetic signals are much weaker for the (110) and negligible for the (111) orientations.   
We attribute the magnetic order to
the disordered graphite layer produced by the laser at the diamond surface. The ferromagnetic signal vanished after chemical etching or after
moderate temperature annealing.  
The obtained results indicate that laser treatment of diamond may pave the way to create ferromagnetic
spots at its surface.
\end{abstract}
\maketitle

\section{Introduction}
\label{introduction}

 Since  the first studies on the magnetic order found in pure graphite-based samples were reported, see \cite{func16} and Refs. therein, the possibility of having
magnetic order in other carbon-based compounds  at room temperature and without 
doping with magnetic ions  attracted the interest of the community. 
In case of pure diamond, Talapatra et al.\cite{tala05} reported the existence of ferromagnetic hysteresis
 at room temperature in the magnetization  of  nanograins of diamond after nitrogen and carbon irradiation. This interesting result 
 was ascribed to structural modification or defects produced by the irradiation, a clear case of the phenomenon called 
 defect-induced magnetism (DIM).
In contrast to the $^{12}$C implantation, a higher value of the magnetization at saturation was 
obtained after $^{15}$N implantation, which was interpreted as due to the extra contribution of N-related centers in the diamond 
crystalline structure \cite{tala05}. Superconducting (with a transition temperature of $T_c \sim 3~$K)  and
ferromagnetic (Curie temperature $T_C > 400~$K) states  were found in hydrogenated boron-doped nanodiamond films by Zhang et al.\cite{zha17}.
Narayan and Bhaumik reported ferromagnetic states
after quenching carbon from an undercooled state using nanosecond laser pulses \cite{nar15}. The  observed magnetic state at room
temperature, which depended on the energy and number of laser pulses,  
was attributed to a mixture of sp$^2$-sp$^3$ bonds in  the nanostructure of the diamond samples. Theoretical work studied the
possibility of ferromagnetism in diamond taking into account disorder and certain doping \cite{ken04}. \color{black} As in other  carbon-based structures
 \cite{barbara04,kob06,ma05,sai05,lee05,ohldagnjp,mak11,fri10}, 
 H-atoms or H$^+$ in the diamond lattice might also trigger a finite magnetic moment, although the influence of its position in the diamond lattice on
 the magnetism has to be
 still clarified \cite{ken04}.

 In this work, in contrast to the above mentioned studies about triggering magnetic order in the diamond structure,  we are mainly interested 
to study the possible development of  magnetic order through a graphitization of the diamond surface via laser pulses. 
Several experimental and theoretical studies on the origin of ferromagnetism in graphite without magnetic impurities have been published over the last 20 years, 
for reviews see Refs.~\cite{chap1,chap2,chap3}. 
With a density of lattice defects or hydrogen between 5\% and 10\%,  graphite can be magnetically ordered with a strong spin polarized
valence band, which affects the polarization of the barely occupied conduction band (graphite is a narrow-band-gap semiconductor \cite{gar12,ari21}). 

It has been known for almost 20 years, see \cite{wan00} and Refs. therein, that the 
surface of pure diamond can be graphitized via laser pulses. 
The heating of the diamond under the influence of laser radiation leads to graphitization, ablation and burn of carbon material \cite{jes99,wan00,tak03}.
The characteristics of the graphite structure at
the surface of the diamond sample (e.g. defects density, crystal orientation, etc.) partially depend 
 on the crystal orientation and length of the laser pulse \cite{wan00,tak03}.   
In a recently published work, the effect of the cutting fluence (of a 532~nm wavelength laser with a pulse duration of 40~ns and
a spot diameter of $40~\mu$m) on CVD diamond surface was investigated with Raman and transmission electron microscopy (TEM) \cite{mou20}.
The authors found that the subsurface of the diamond samples shows a mixture of graphite and amorphous
carbon and that the thickness of the graphite layer decreased with laser fluence. Systematic studies on this topic have been published 
earlier \cite{her15}. However, no magnetic characterization of
the produced graphite/amorphous carbon layers was reported.  
In this study, we have used laser pulses of 532~nm wavelength, 300~J/cm$^2$ energy density in $15~\mu$m focus spot and 200~ns pulse duration at
15~kHz to produce a  graphitic-like layer at the surface of several diamond samples and studied their magnetic properties. \color{black}

\section{Samples and Methods}
\subsection{Laser cutting and after-cutting processes of diamond crystals}
\label{spre}
A single crystal of diamond is glued to the base surface of a mandrel, so that a large face is orientated perpendicular to 
the axis of the mandrel. Next, the mandrel (with the diamond crystal) is fixed in a device for the precise 
positioning  in the laser-cut system. The system is equipped with a video camera that allows to adjust the face 
of the crystal to be cut along the axis of the laser beam to achieve the shortest laser cut length.
The marking of the cut line on the selected  face of the diamond single crystal is carried out 
on the computer monitor with the help of optical devices. 


Before cutting, the laser beam was focused on the surface of the diamond at the level of the upper point of the cut. Then the laser beam 
was moved along the cut line, where the material was burned on the surface of the diamond with a width 
nearly  the diameter of the laser beam focus. Thus, the working pass was performed at a certain depth in the crystal. 
After leaving the diamond sample, the laser beam was moved by a step (specified in the software) in the transverse direction and then moved in the opposite direction performing the next working pass.  By selecting the wavelength, power, and the duration of the laser radiation pulses, we can  control the volume of material removal.

\begin{figure}[h]
\begin{center}$
\begin{array}{cc}
\includegraphics[width=2.5in]{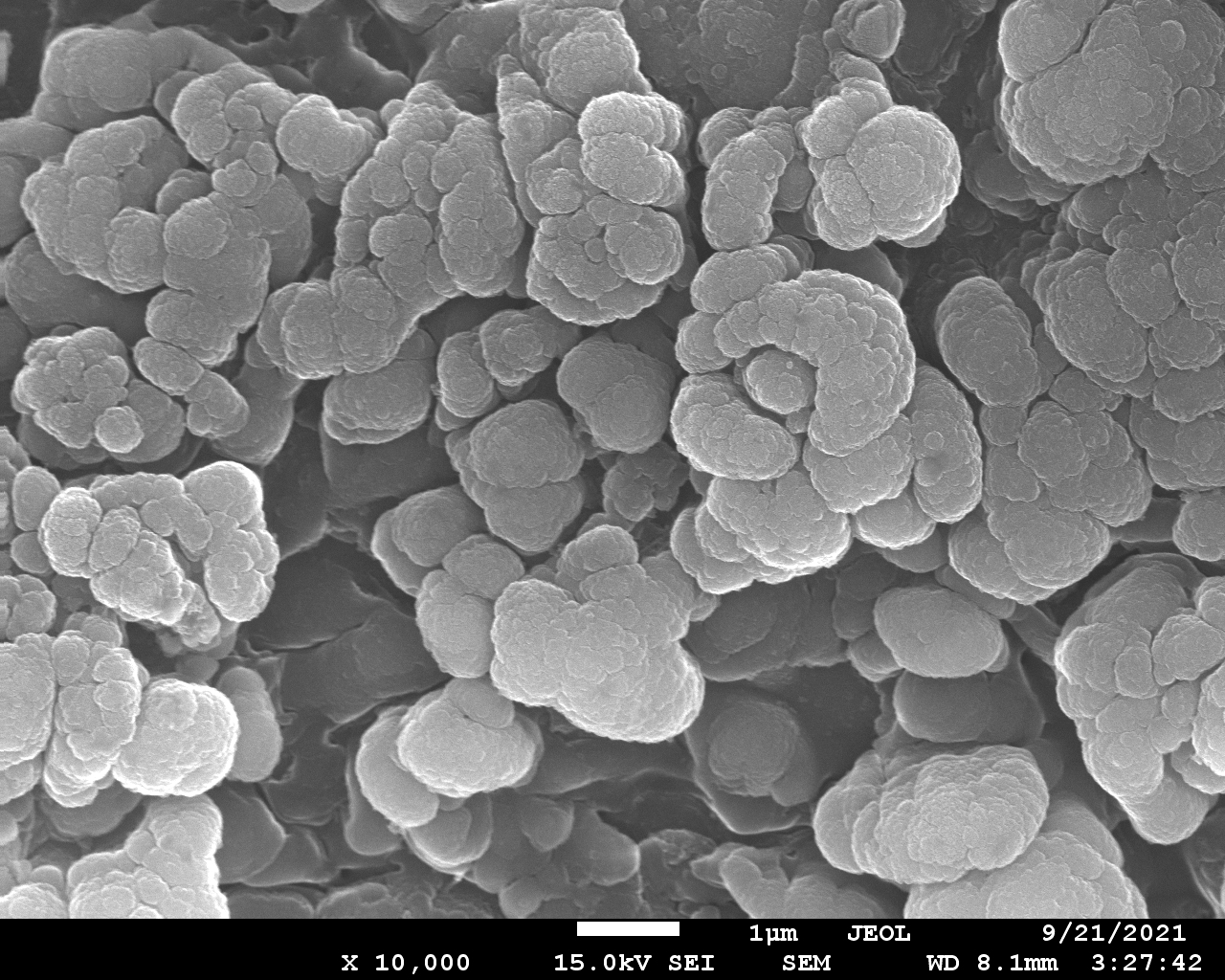} \\
\includegraphics[width=2.5in]{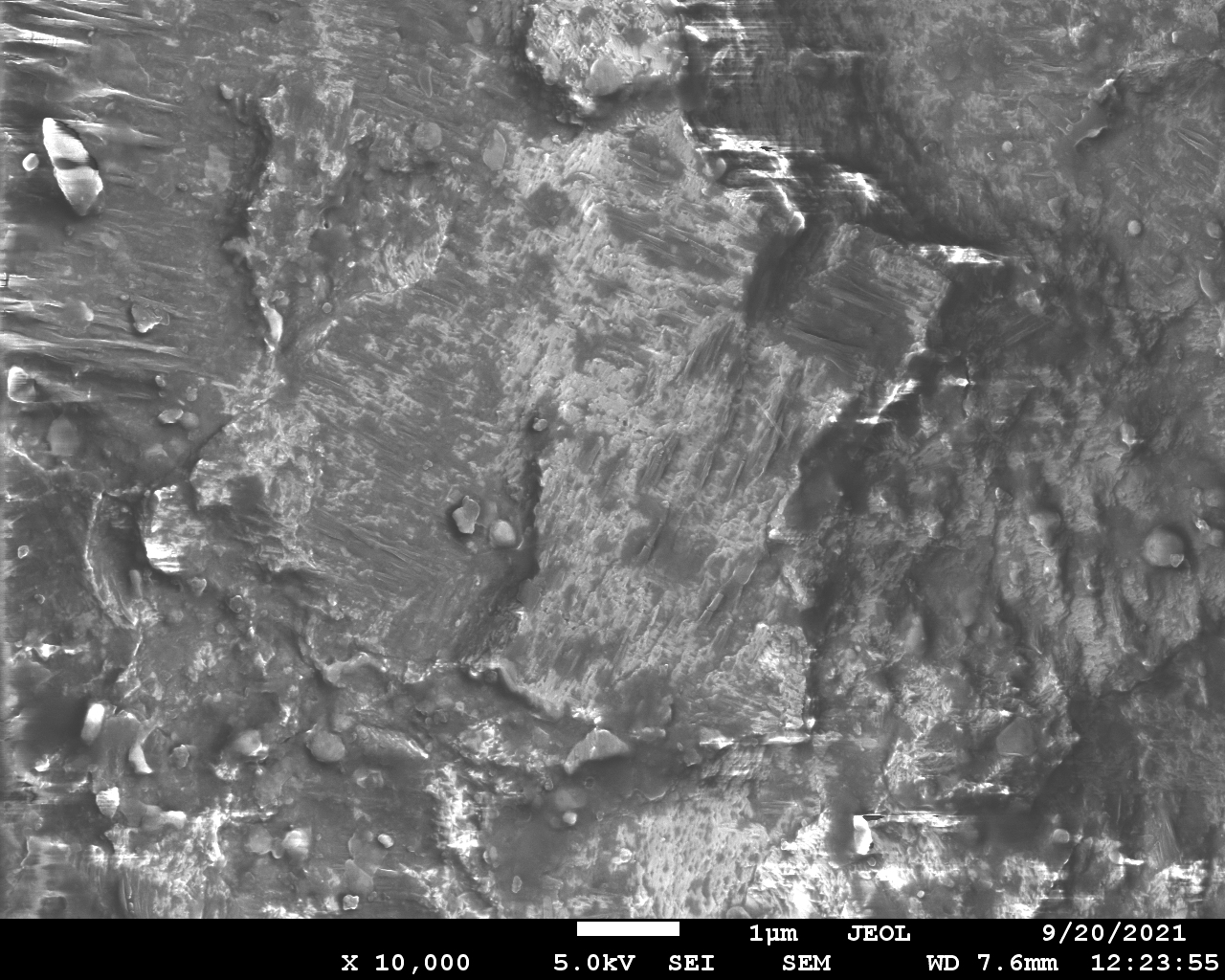}
\end{array}$
\end{center}
\caption{Scanning electron microscope images of the surface of a CVD diamond sample after laser cut (upper image) and after laser polishing (lowert image)}
\label{sem}
\end{figure}

After performing several working steps, the laser beam reached half depth of the crystal to be cut. 
The shape of the cutting groove is wedge due to a conical shape of the focused laser beam. 
Thus, to reduce the amount of the ablated (burned) material, the mandrel with crystal is rotated $180^\circ$ and the cutting process 
continued from the back part of the crystal to cut its rest half of the thickness. 
To reduce the loss of diamond material during cutting, the cutting angle and the cutting width at the output are  as small as possible. On the other hand, the laser beam must have a relatively large convergence angle  to ensure high-quality focusing. With a significant reduction of the cutting angle, most of the laser power is reflected, which slows down the cutting process. Therefore, in order to optimize this process, we do a counter cut, which in turn helps to reduce the cutting depth and material losses.
A SEM image of the obtained surface can be seen in Fig.~\ref{sem}(left).

After cutting,  a polishing  on the cut face was performed. 
The laser beam with the specified parameters was focused directly on the surface of the sample 
in order to burn off part of the surface material produced by the cutting.
A SEM image of the obtained surface can be seen in Fig.~\ref{sem}(right).


\color{black} In order to remove the graphitic-like nanometers thick ferromagnetic surface region formed after laser treatment we used two methods: (1) Chemical etching of 
the laser cut samples   
 with a mixture of 30~mL concentrated sulphuric acid (H$_2$SO$_4$), 10~mL fuming 
 salpetric acid (HNO$_3$) and 10~mL 70 vol\%
 perchloric acid (HClO$_4$). This mixture was heated at 120~C for 4~h. under reflux. 
 After cooling to room temperature, the acids were decanted and the diamond was intensively washed with distilled water and dried with nitrogen gas. 
 In comparison to literature \cite{sas12,pol18}, we have applied a modified etching procedure at higher temperatures with a mixture of strong oxidizing acids to remove  the 
 graphite residues derived from the laser treatment. 
With this procedure  we estimate that the  
disordered graphite thickness that the etching process removes should be at least $\sim 20~$nm. Certainly, not all this
thickness might be magnetically ordered. From a comparison between the  magnetization at saturation values of  
ferromagnetic graphite \cite{ohldagnjp,chap3} and our laser-treated samples, we estimate that the ferromagnetic thickness 
 should be $\lesssim 20~$nm, see Section~\ref{magn}.  
 (2)  The other method we used is annealing the diamond sample in air at  temperatures $T \leq 650$~C for a couple of hours.  
\color{black} 

\subsection{Samples characteristics}
\subsubsection{Natural diamond samples}

\color{black} Table~\ref{nc} shows several characteristics of the natural diamond samples like the orientation of the laser cut surface as well as 
the total nitrogen concentration N and the following nitrogen-related defect concentrations: -A:   a neutral nearest-neighbor pair of nitrogen atoms substituting the carbon atoms. 
-B:  a carbon vacancy surrounded by four nitrogen atoms substituting the corresponding carbon atoms. -C: electrically neutral single substitutional nitrogen atoms in the diamond lattice, sometimes called also P1-center, see, e.g., \cite{bab08,woo84}. \color{black}

\begin{table}[] 
\caption{Size, orientation, total Nitrogen concentration (N) and of A, B, and C centres of the natural diamond samples }
\begin{tabular}{cccccc}
\toprule
\textbf{Name}	& \textbf{Orientation}	& \textbf{Mass} & \textbf{Cut area} & \textbf{N} & \textbf{N-centres}\\
                          &                                 & mg  & mm$^2$ & ppm & ppm (A,B,C)\\
354		& (100)			& 96.0 & $45 \pm 3$ & 3170 & 195, 681, 56\\
356		& (100)			& 106.0 & $15 \pm 3$ & 1400 & 522, 78, 46\\
540		& (100)			& 96.0 & $25 \pm 3$ & 750 & 240, 66, 9\\
164		& (111)			& 38.6 & $20 \pm 3$ & 3900 & 605, 656, 81\\
384		& (111)			& 120.0 & $40 \pm 3$ &2000 & 696, 132, 57\\
\end{tabular}\label{nc}
\end{table}

\subsubsection{CVD diamond samples}
As we will demonstrate below in this paper, the orientation of the laser cut surface plays a main role to produce the
robust ferromagnetic nanometer thick surface region at room temperature. To support  the results obtained from the natural
diamond crystals, we have cut 6 CVD diamond samples at three orientations (100), (110) and (111),
see Table~\ref{cvdt}. These samples have a  total concentration of magnetic impurities below 2~ppm and
a much lower N concentration ($\lesssim 10~$ppm)  than  the natural diamond samples. We measured the magnetic response of the CVD  
samples after the first cut (state "a") and after polishing  the cut surface with the  laser beam (state "b"). There are basically no  differences in the
magnetic behavior between the states "a" and "b", expect  that after removing part of the cut surface by polishing the
ferromagnetic signal at saturation becomes smaller. 

\begin{table}[] 
\caption{Sample name, surface orientation, laser treatment, mass, and cut area of the CVD diamond samples.}
\begin{tabular}{ccccc}
\toprule
\textbf{Name}	&\textbf{Orientation} & \textbf{Laser treatment}	&\textbf{Mass}&  \textbf{Cut area}  \\
                  &        &                               &mg  &  mm$^2$  \\
1a		& (100)	&cut		&32.5 & $14.4 \pm 0.4$  \\
1b		& (100)	&polish		&33.6 &  $14.4 \pm 0.4$  \\
2a		& (110)	&cut		& 35.6&  $5.7 \pm 0.2$ \\
2b		& (110)	&polish		&33.3 & $5.7 \pm 0.2$  \\
3a		& (111)	&cut		&39.5 & $7 \pm 0.2$  \\
3b		& (111)	&polish		& 65.7& $6.7 \pm 0.2$   \\
\end{tabular}\label{cvdt}
\end{table}

\subsection{Methods: PIXE, Raman and SQUID characterization} 

The quantitative characterization of the main magnetic impurities (Fe, Co and Ni) was done using 
Particle Induced X-ray Emission (PIXE) with protons. The  parameters were,
proton energy: 	2.0~MeV, 
	current:	2.5~nA,
	slit settings: 	Object/Aperture: 300~$\mu$m/300~$\mu$m, and 
a	beam focus of	3~$\mu$m.   Protons of 2 MeV kinetic energy have a penetration range in diamond of about 25~$\mu$m. The X-ray production cross section decreases as the protons slow down inside the sample. Additionally, the contribution of X-rays to the detectable analytical signal decreases with the depth. However,  this effect for X-rays from Fe, Co, Ni is less important due to the rather short range of 2~MeV protons in carbon matrix. 

Confocal Raman measurements were performed with the WiTec Alpha 300 System. A Laser wavelength of 532~nm (UHT S 300) was selected. 
A 50X objective (Zeiss, Germany) with a numerical aperture of 0.8, a laser power at the sample surface of ca. 35~mW and  
a 1800 grating on the CCD detector with spectral resolution of ca. 0.8~cm$^{-1}$ were used.

The measurements of the magnetic moment of the diamond samples have been done with a Superconducting Quantum Interferometer Device (SQUID) 
 from Quantum Design. 
Magnetic field loops and temperature hysteresis were obtained after demagnetizing the samples at 380~K. The time between two consecutive
measurements at different fields or temperatures was 5~min or longer with similar results. No
time dependence was detected within experimental resolution.

\section{Results}
\label{res}

\subsection{Magnetic impurities measurements}
From the  characterization of the impurities with PIXE we conclude that
the maximum magnetic impurity concentration was 2.6~ppm of Fe in sample 164. The sample 354, which shows  the largest magnetization
at saturation, has a total concentration of magnetic impurities below 0.5~ppm, see Table~2. 

As example, let us estimate how large would be the contribution 
of 0.17~ppm Fe in sample 354  to the magnetization at saturation. 
We assume that this small concentration of Fe or most likely magnetite, 
Fe$_3$O$_4$, behaves as a bulk ferromagnet with a saturation magnetization of
about 100~emu/g. The measured concentration of Fe in sample 354 would imply
a ferromagnetic total mass of 76~ng, which in the unrealistic largest 
case could contribute with
a magnetic moment of $4.8~\mu$emu. This value is 4.2 times 
smaller than the magnetic moment 
at saturation measured at 300~K (see Section \ref{mnd} below).

\begin{table}[] 
\caption{Main magnetic impurities content (in $\mu$g/g) measured by Particle Induced X-ray Emission (PIXE) of the natural diamond samples. The selected areas for
the measurements are included. The spot area was $9~\mu$m$^2$. MDL: minimum detectable limit.}
\begin{tabular}{cccp{0.95cm}cccc}
\toprule
\textbf{Sample}	 &\textbf{Area}	& &\textbf{Concentration}  & & & \textbf{MDL} &\\
	& 	& \textbf{Fe} & \textbf{Co} & \textbf{Ni} & \textbf{Fe} & \textbf{Co} & \textbf{Ni}\\
354		& $(0.5~$mm)$^2$			& 0.17 & 0.046 & 0.16 & 0.03 & 0.03 & 0.03\\
                 & $(160~\mu$m)$^2$ & 0.08 & $<0.07$ & $< 0.07$ & 0.05 & 0.04 & 0.05\\
356		& spot			& 1.99 & - & 0.12 & 0.07 & 0.04 & 0.05\\                 
379		& $(0.5~$mm)$^2$			& 1.85 & $<0.03$ & 0.22 & 0.04 & 0.03 & 0.04\\
                 & $(240 \times 100)~\mu$m$^2$ & 0.72 & - & $< 0.12$ & 0.1 & 0.1 & 0.1\\
                 540		& spot			& 0.57 & - & 0.11 & 0.06 & 0.04 & 0.04\\
164		& $(0.5~$mm)$^2$			& 2.6 & 0.038 & 0.098 & 0.04 & 0.03 & 0.03\\
                 & $(75~\mu$m)$^2$ & 0.32 & - & $< 0.17$ & 0.2 & 0.2 & 0.2\\
                 384		& (1~mm)$^2$			& 0.84 & 0.28 & 0.35 & 0.08 & 0.08 & 0.07\\
                 & $(400~\mu$m)$^2$ & 0.8 & $<0.18$ & $ 0.17$ & 0.2 & 0.2 & 0.2\\
\end{tabular}\label{nci}
\end{table}

\begin{figure}
\includegraphics[width=8 cm]{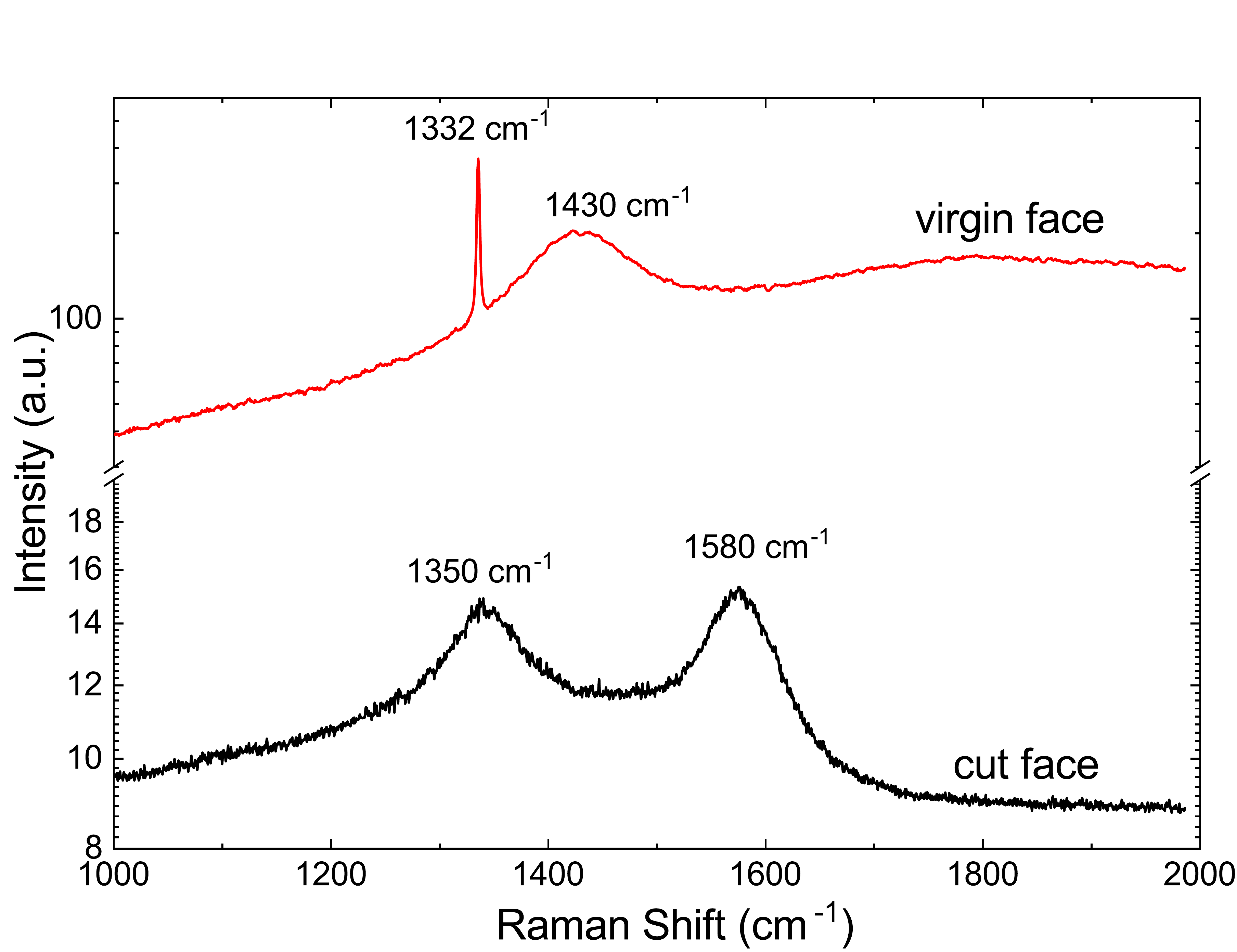}
\caption{Raman Spectra obtained at room temperature for the virgin and laser cut faces   of  the CVD sample \#1b. }
\label{raman}
\end{figure}

\subsection{Raman}
\label{ram}
Raman measurements were done on the "virgin", i.e. a region of the same sample without any laser treatment, and on the laser cut surfaces of all samples. 
As example, we show in  Fig.~\ref{raman} the results of the CVD sample \#1b. 
Whereas the virgen surfaces of the samples show a sharp absorption peak
 at 1332~cm$^{-1}$ corresponding to pure diamond (first order Raman), the laser-cut surfaces show disordered graphite-like 
 peaks due to the G-band (1580~cm$^{-1}$) and D-band (1350~cm$^{-1}$), see Fig.~\ref{raman}. In case of the virgin  surfaces,
also the peak at 1430~cm$^{-1}$, observed in CVD samples \cite{sti96,bad97,zai01},  is clearly observed.

We observed some differences in the Raman patterns between the cut surfaces of the CVD "a" and "b" samples: whereas in the "b" samples
only Raman peaks corresponding to disordered graphite were observed, the "a" cut surfaces also showed  a weak signal coming from the diamond main Raman peak.  
Apparently, the polishing procedure  transforms the rest of diamond-like regions left after the first laser treatment, leaving only disordered graphite regions. 
Within the experimentally observed broadening of the Raman peaks, it is not possible  to recognize 
systematic differences between the Raman spectra of the different cut surface orientations. Clearly, Raman characterization
helps to identify the presence of graphite-like regions (after the laser treatment) but it does  not provide clear
hints for   the presence of certain defects that can be correlated with the magnetic response.

\subsection{Magnetization measurements}
\label{magn}
\subsubsection{Natural diamond samples}
\label{mnd}
\begin{figure}
\includegraphics[width=10 cm]{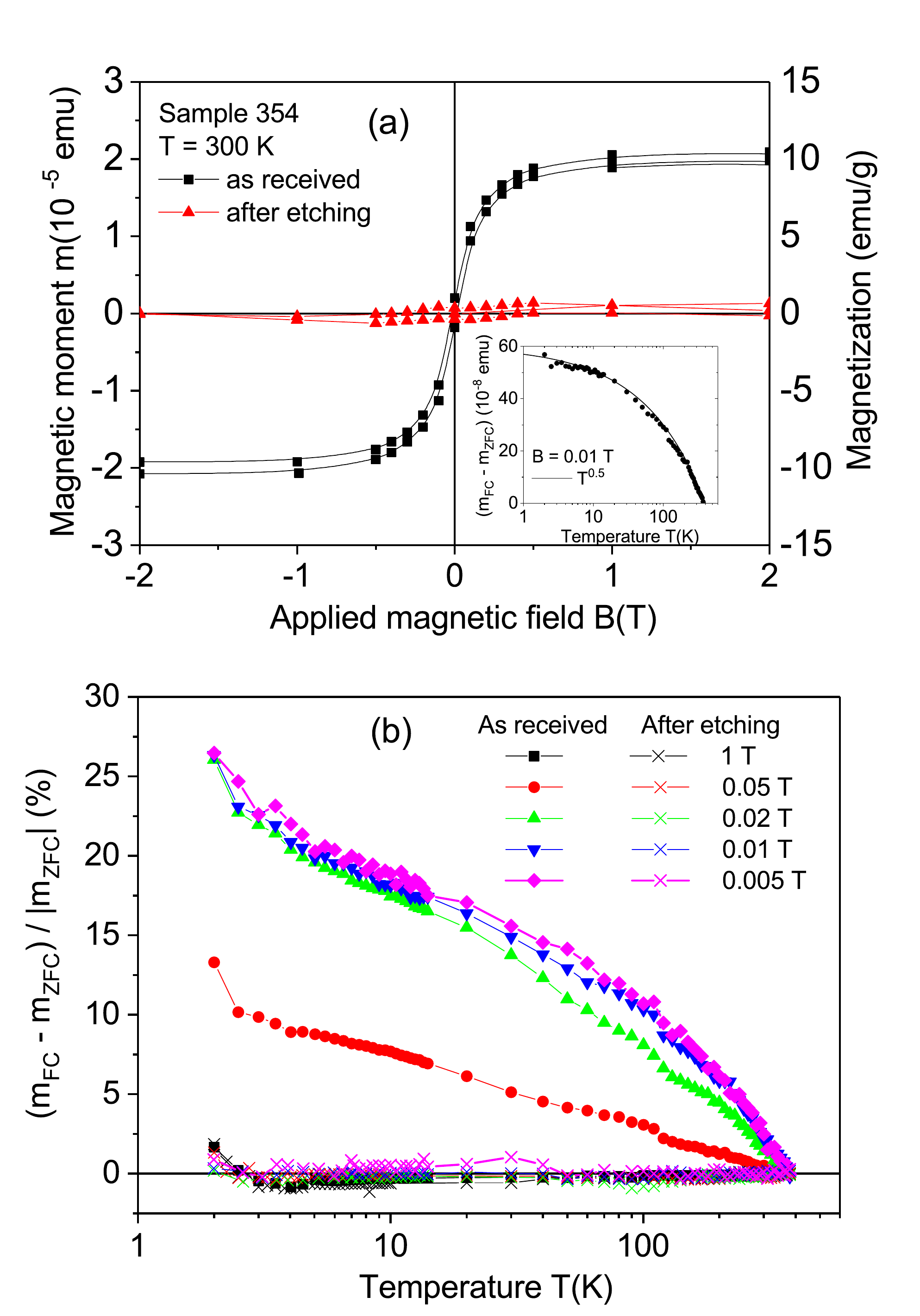}
\caption{(a) Field hysteresis loops at 300~K of sample 354 in the as-received  state and after chemical 
etching. The same linear diamagnetic background was subtracted
from the measured curves. The inset shows the temperature dependence of the 
difference in the magnetic moment measured in the field cooled (FC) and zero-field cooled (ZFC) states measured at a fixed field
of 0.01~T. The continuous line follows the equation $m_{\text FC}(T) - m_{\text ZFC}(T) = 10^{-8} (60 - 3T^{0.5})$ (emu). (b) Temperature
dependence of the relative difference $100 [m_{\text FC}(T) - m_{\text ZFC}(T)] / |m_{\text ZFC}(T)|$ before and after etching.}
\label{1}
\end{figure}  

\begin{figure}[]
\includegraphics[width=8cm]{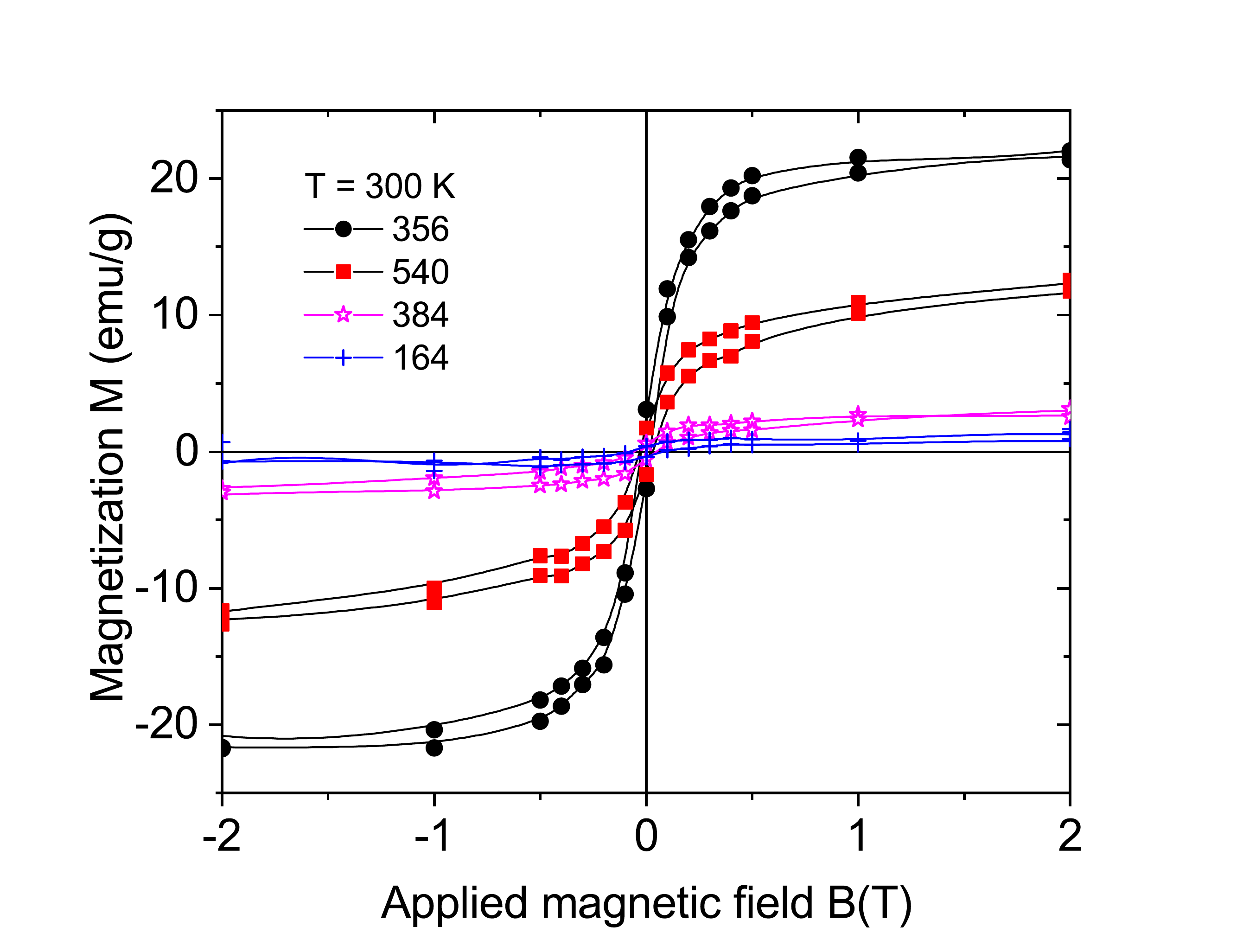}
\caption{Field hysteresis loops at 300~K of four natural diamond samples   ((100) surface: 356 and 540,  (111) surface: 164 and 384). 
The magnetization values were calculated assuming a 
ferromagnetic mass given by a 20~nm thick region at the cut surfaces of the samples.}
\label{2}
\end{figure}
   
\begin{figure}[]
\includegraphics[width=7 cm]{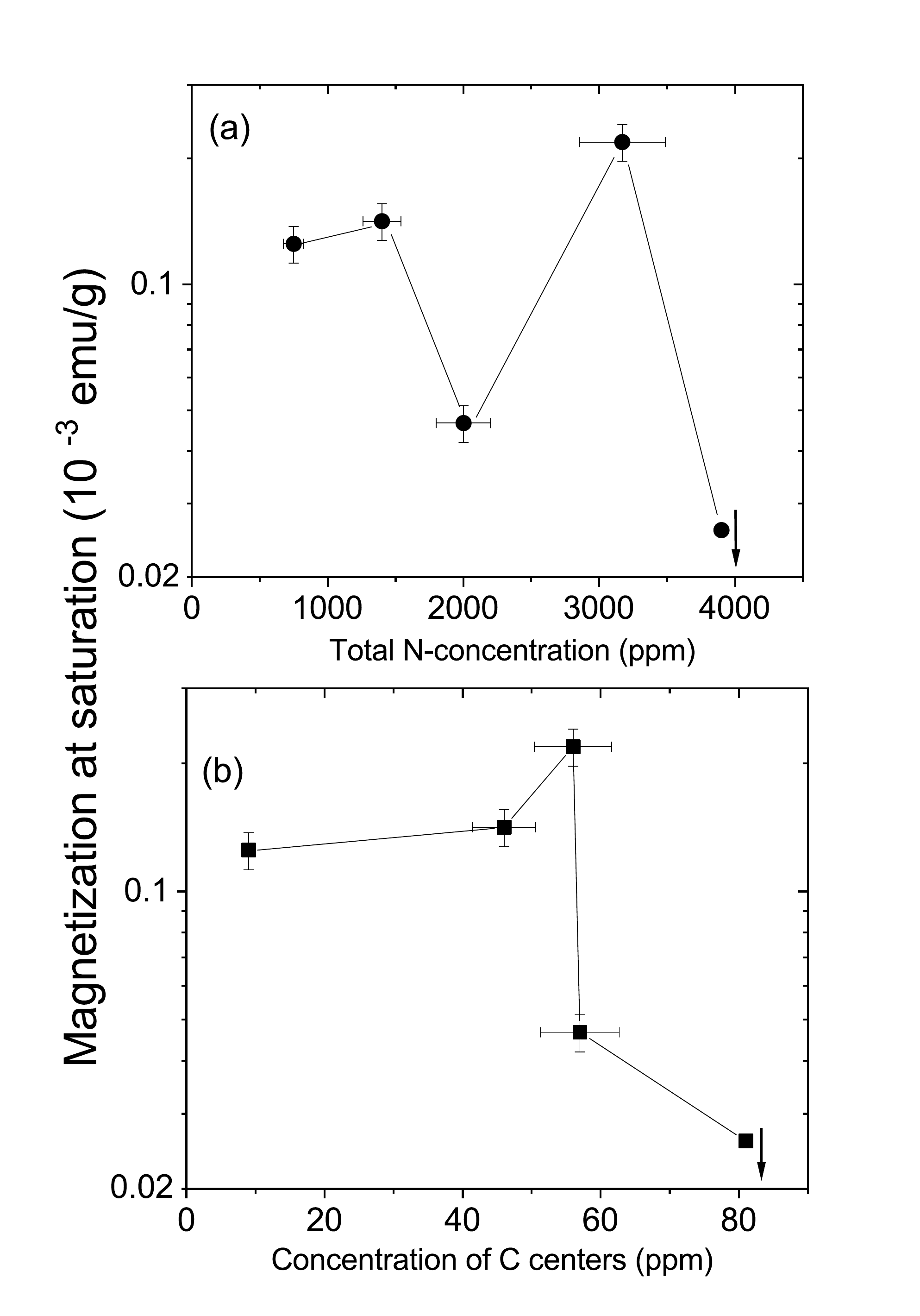}
\caption{Magnetization at saturation at 300~K  of the natural crystals vs. the total concentration of nitrogen (a) and of C centers (b). Because nitrogen is in all the 
sample volume, the magnetization has been calculated taking into account the total mass of each sample. The downarrow at the  last point to the right means that the
value of magnetization for that sample is below the limit of the y-axis. }
\label{MN}
\end{figure}

 \color{black} With the magnetic impurity concentration of our samples, 
the natural diamond crystals in the virgin state (before any laser treatment)  do not show any sign of magnetic order 
at 300~K within the resolution of our SQUID magnetometer ($\sim 5 \times 10^{-8}$~emu at an 
applied field of 1~T).  Figure~\ref{1}(a) shows the field hysteresis loops of sample 354 at 300~K   
before (as-received) and after chemical etching,  within $\pm 2$~T
field range. The same diamagnetic linear contribution was subtracted from both data sets. Before chemical etching, the sample shows 
a clear ferromagnetic response. 
In the inset of Fig.~\ref{1}(a) we plot the
difference between the field cooled (FC) and zero field cooled (ZFC) states at 0.01~T applied field. This difference follows a temperature dependence similar to
that found in irradiated graphite \cite{jems08,chap3}. This similarity and  the Raman results, see Section~\ref{ram},  
indicate that the disordered 
graphite layer produced by the laser treatment should be at the origin of   the observed ferromagnetic response. 
As a  proof for this assumption, the same sample was treated chemically to remove the disordered 
graphitic layer.  The reduction of the ferromagnetic response observed in the field hysteresis loop of  Fig.~\ref{1}(a)
after chemical etching, clearly indicates that the ferromagnetic behavior is related to the graphitic-like layer produced by the
laser treatment.  \color{black}

To further demonstrate the large difference in the ferromagnetic response between the cut sample before and after etching, 
the difference between the FC and ZFC states relative to the value in the ZFC state given by $100 [m_{\text FC}(T) - m_{\text ZFC}(T)] / |m_{\text ZFC}(T)|$
at different applied magnetic fields is shown in Fig.~\ref{1}(b). In this figure we recognize that whereas this relative difference
reaches $\sim 25$\% \color{black}(of $|m_{\text ZFC}(T)|$)  \color{black}  at low temperatures and at fields $\le 0.02~$T in the as-received sample, 
it remains below 1~\% in the whole temperature range and applied fields 
after etching the sample. \color{black} These results indicate further that those signals are related to the surface near graphitic-like region.

 With the estimate ferromagnetic  thickness of $\sim 20~$nm and
the measured area of the laser cut surface, we obtain  
 a  magnetization (right y-axis in Fig.~\ref{1}(a))  at saturation of 10~emu/g. 
A comparison with the values of the magnetization at saturation obtained for ferromagnetic graphite \cite{ohldagnjp,chap3}, we note that this ferromagnetic 
thickness should be of the order or even smaller.  \color{black} 



Figure~\ref{2} shows the field hysteresis loops of four natural diamond samples with cut areas with orientation (100) (356 and 540, similar to sample 354, see Table~\ref{nc}) and
with (111) orientation (164 and 384) at 300~K. Taking into account  the cut area and assuming the same ferromagnetic thickness, we recognize
that the ferromagnetic signals are clearly smaller for the (111) cut surface samples.  This difference is not related to large differences in the
assumed ferromagnetic mass because the cut surfaces are similar or their difference shifts the estimate value of magnetization in the opposite direction, 
see Table~\ref{nc}. \color{black} This result indicates that the diamond crystalline structure and the laser cut direction relative to its structure play an important role 
to trigger  the ferromagnetic order in the graphitic-like surface layer. The results obtained from the CVD samples support this conclusion, see Section~\ref{cvd-s}. 

 Nitrogen doping with the concentrations measured in our samples, or lower, see Table~\ref{nc},  does not trigger 
ferromagnetic order at room temperature.The ferromagnetic signal is not related to the total N-concentration or to the concentration of three defect centers  one finds in 
bulk N-doped diamond  (A, B, C), see Figs.~\ref{MN}(a) and (b). In   Figs.~\ref{MN}(a) and (b) and due to the fact that these centers are distributed all over the samples,
the shown magnetization values  were obtained taking into account the whole sample mass. 
On the other hand,  we note that N-related C-centers are at the origin of the clear hysteretic behavior in field and temperature
observed below 50~K  \cite{bardia,set21}. 
\color{black}

\begin{figure}[]
\begin{center}
\includegraphics[width=9.0 cm]{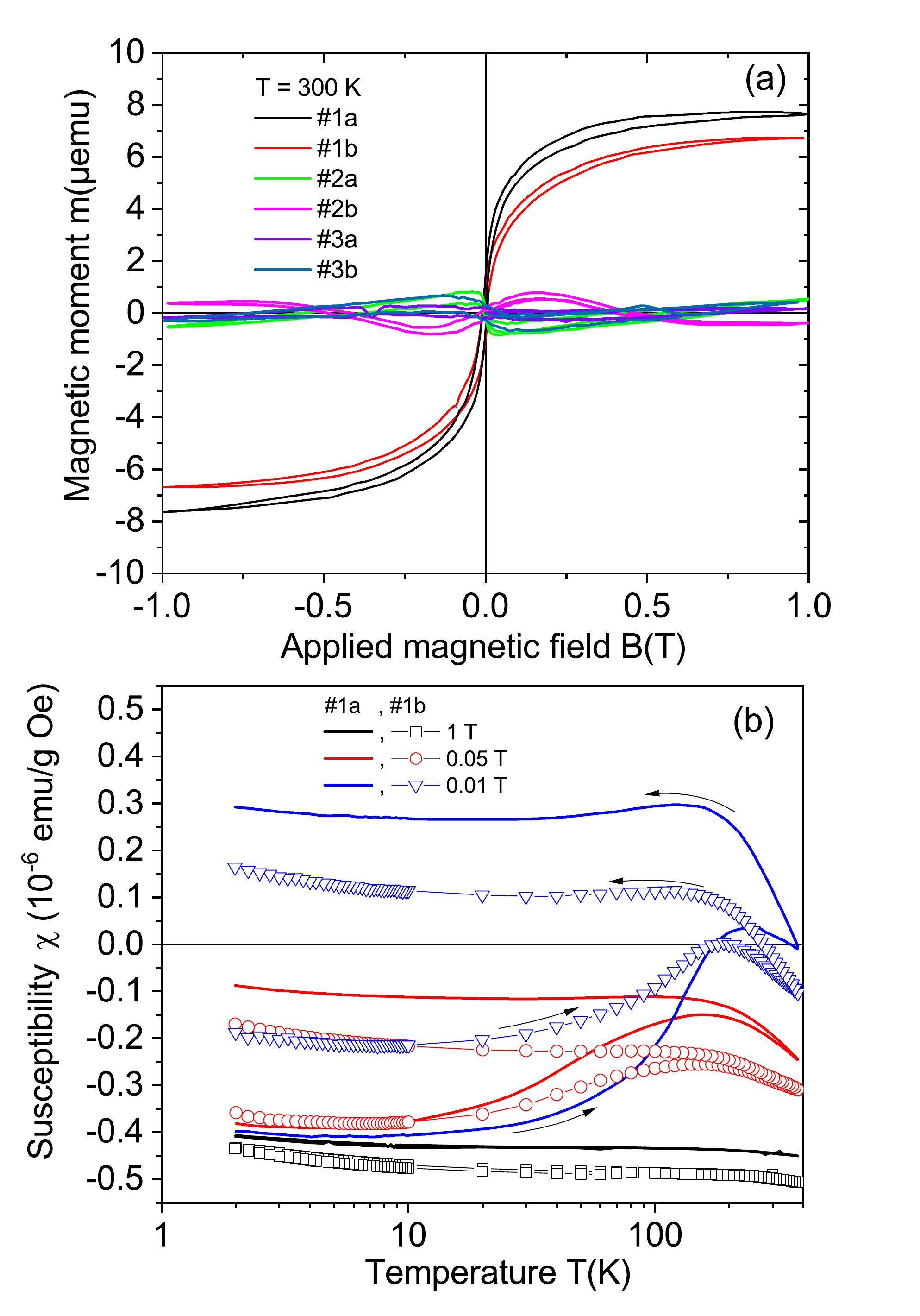}
\caption{(a) Field hysteresis loops of the magnetic moment of all  CVD samples  at 300~K. The diamagnetic background was subtracted from the 
measured data.  We note that the field loops obtained for some of the CVD samples are influenced by the hysteresis of the superconducting
 solenoid. This is a systematic error of the SQUID system that can influence the  hysteresis loops. 
 (b) The susceptibility as a function of temperature at constant fields in the zero-field cooled (measurement by warming the sample
 after applying the corresponding fields at the lowest temperature) and field-cooled (measurement by cooling the sample at the same field) 
states of the two samples (\#1a: continuous lines, \#1b: symbols). No background
was subtracted from the data. The mass used to calculate the susceptibility was the total sample mass, which is similar in both samples. }
\label{cvd}
\end{center}
\end{figure}

\subsubsection{CVD diamond samples}
\label{cvd-s}

 \begin{figure}[]
\includegraphics[width=9 cm]{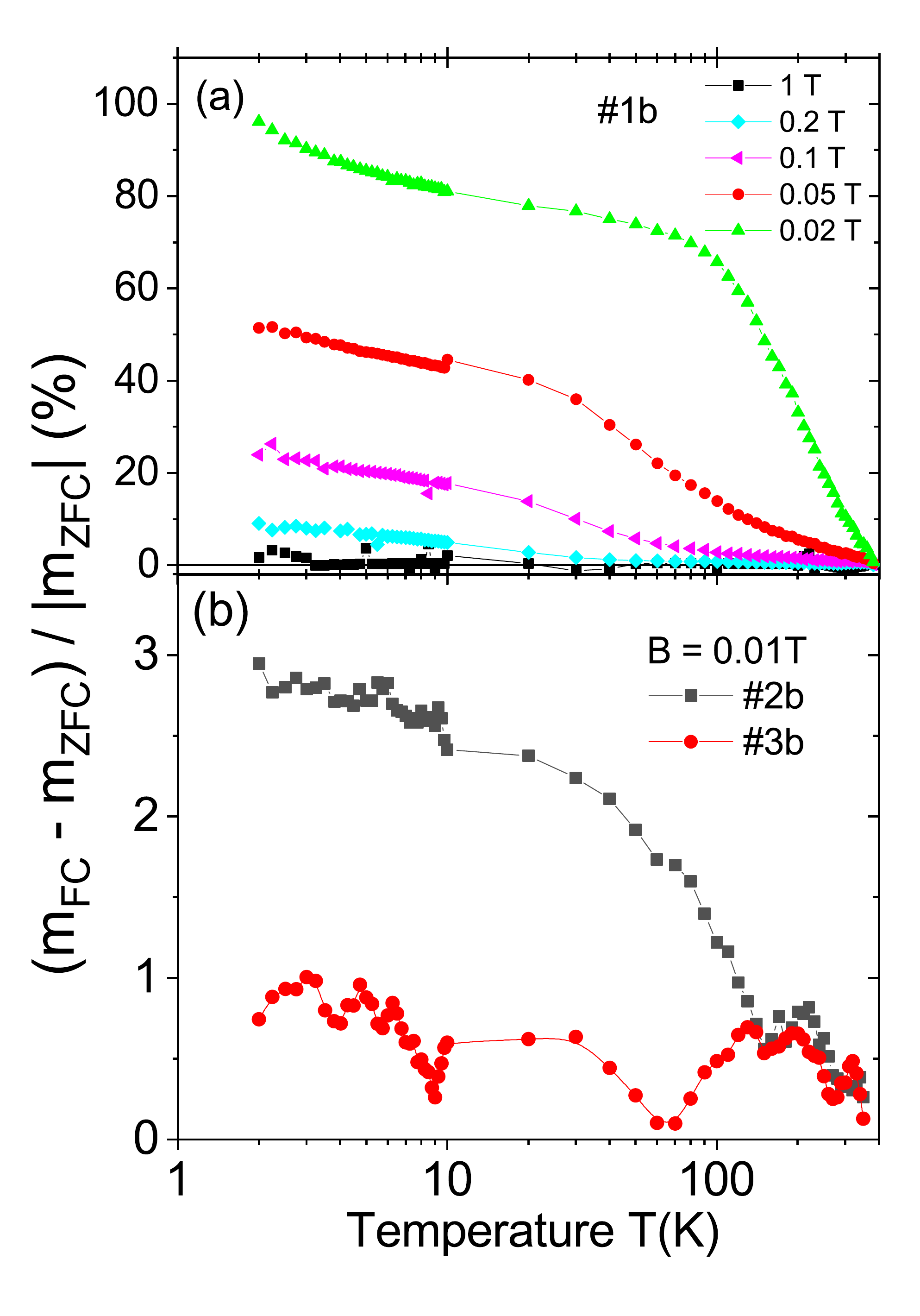}
\caption{Temperature
dependence of the relative difference $100 [m_{\text FC}(T) - m_{\text ZFC}(T)] / |m_{\text ZFC}(T)|$ for the CVD samples: 
(a) \#1b at five different applied fields and (b) 
 \#2b and \#3b at a field of 0.01~T.}
\label{cvd2}
\end{figure}

 Figure~\ref{cvd}(a) shows the field hysteresis loops measured at 300~K of all CVD samples, see Table~\ref{cvdt}, 
 after subtracting the  linear diamagnetic
 background.  \color{black} The results  indicate a ferromagnetic behavior with a coercive fields of the order of 80~Oe for  samples \#1a and \#1b.
 The magnetic moment at saturation is much smaller for the samples with other crystal orientations.
 Taking into account the
 volume of the cut surface or the total mass, the obtained ferromagnetic 
 magnetization of  samples \#1a and \#1b is always larger than that of the other CVD samples, 
 supporting the orientational dependence of the ferromagnetic signals of the laser cut surface observed in the natural diamond crystals. 
 As in the natural diamond crystals, the samples cut with orientation other than the (100) show a much 
smaller or negligible ferromagnetic signal,
see Fig.\ref{cvd2}. The relative difference between the FC and ZFC curves is nearly two orders of magnitude larger
for samples with (100) cut surfaces. 
  \color{black}
  
 We observe that the saturation magnetic moment of sample \#1b obtained after laser polishing the cut surface,  is about 10\% smaller than of sample \#1a. 
 The ferromagnetic behavior  is clearly observed in the difference between
ZFC and FC states, as shown by  the temperature dependence of the susceptibility, see Fig.~\ref{cvd}(b).  \color{black}  As expected for a ferromagnetic behavior,
 the difference between
ZFC and FC states as a function of temperature vanishes at high magnetic fields, in agreement with the vanishing of the field hysteresis width 
at high enough fields, see  Fig.~\ref{cvd}(a).  \color{black}

\begin{figure}[]
\includegraphics[width=9 cm]{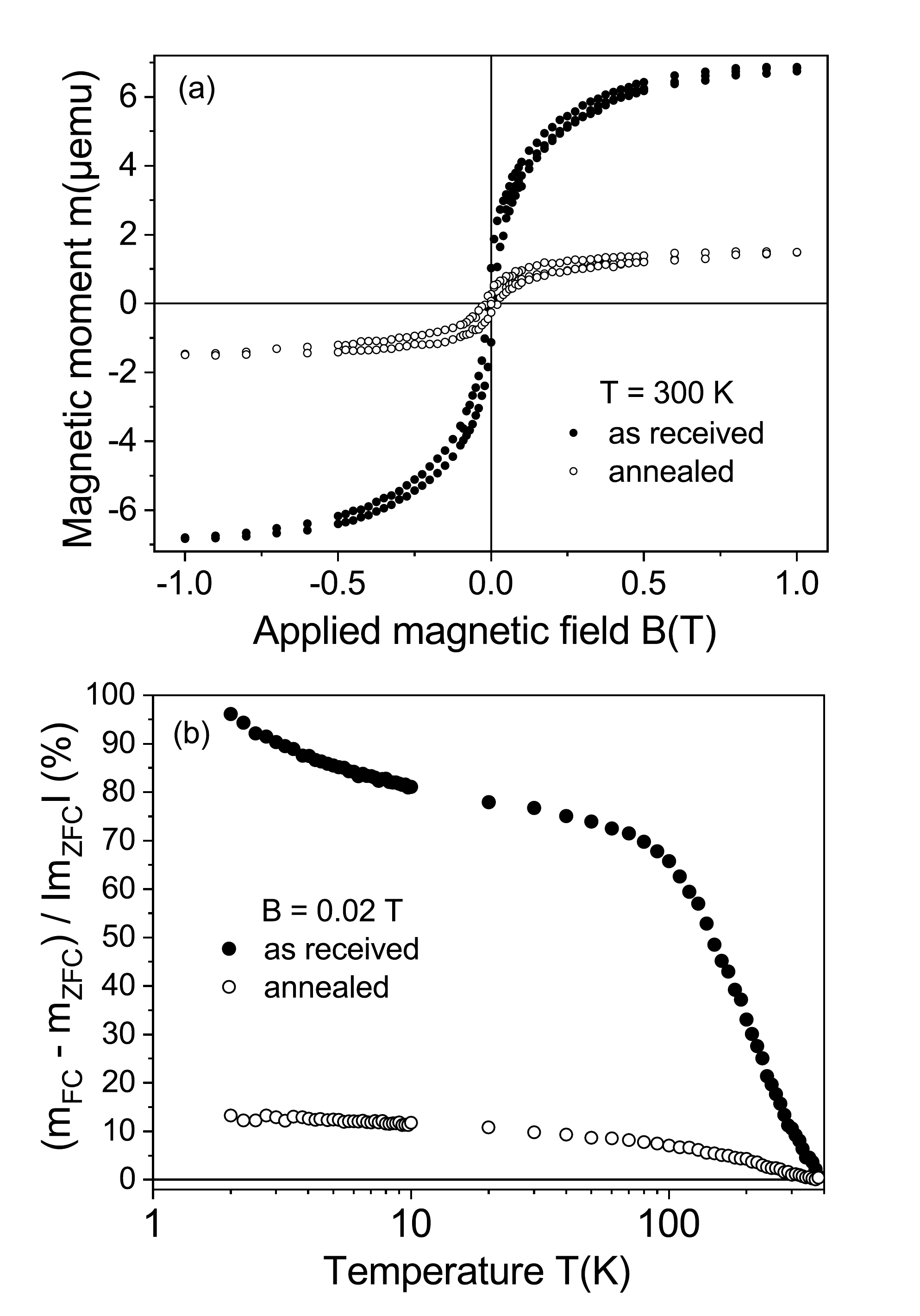}
\caption{Magnetic properties of sample \#1b: (a) Field hysteresis loops at 300~K in the as-received state and after the high-temperature annealing.  (b) Temperature
dependence of the relative difference $100 [m_{\text FC}(T) - m_{\text ZFC}(T)] / |m_{\text ZFC}(T)|$ at a field of 0.02~T, before and after annealing.}
\label{ann}
\end{figure}

It is known that high temperature annealing in air removes any graphitic-like surface regions in diamond. 
Therefore, instead of using chemical etching to remove the graphitic surface of the CVD samples, 
as done in the natural diamond samples (see Fig.~\ref{1}), we have annealed one
of the CVD samples (\#1b) in air. The annealing procedure in air was 1h at 550~C, 1h at 600~C and 0.5h at 650~C. 
Similarly to the result after chemical etching of a natural diamond sample, the ferromagnetic signal strongly decreased 
after annealing, see Fig.~\ref{ann}. All these results clearly indicate that the ferromagnetic signal comes from the disordered graphite
surface region obtained after the laser cut and it is not related to magnetic impurities. 

Before concluding, we would like to remark that the Curie temperature of the ferromagnetic order observed in the
laser treated surfaces of diamond with (100) direction, is clearly larger than 380~K. The  temperature of 380~K is the turning point temperature 
at which  the FC measurement start. For this reason,  the difference $m_{\text FC}(T) - m_{\text ZFC}(T)$
is always zero at the turning point temperature. A rough extrapolation of the observed temperature dependence of the magnetic moment 
to temperatures above 400~K, see for example Figs.~\ref{2}(b) and \ref{cvd}(b), indicates a Curie temperature between 500~K and 750~K, similar
to defect-induced ferromagnetic graphite, see \cite{chap3} and Refs. therein.

\section{Conclusions}

 \color{black}  Independently of the 
 origin of the diamond sample, natural or CVD, we found that under the
selected conditions, the laser pulses
 produced a robust magnetically ordered graphite film at 300~K in  samples  cut   along  the diamond (100) surface orientation. Assuming  a maximum thickness
 of 20~nm for the magnetic layer, the  magnetization value at saturation varies from  $\sim 10~$emu/g to  20~emu/g at 300~K, similar to 
the magnetization values obtained for defect-induced ferromagnetic graphite  \cite{ohldagnjp,chap3}.  
This magnetic order is clearly weaker or absent in the cases of the other two surface orientations.
  \color{black} 
Further focused experimental  characterization but also computer simulations  as in Ref.~\cite{jes99},  are  necessary to 
 find the  lattice defects (e.g., C-vacancies,  sp$^2$-sp$^3$ or  C-H complexes) 
 responsible for the observed ferromagnetism. Laser treatment can in principle be used 
to create localized magnetic spots of small area on a diamond  surface. This phenomenon can be of interest not only for memory
devices but also for other rather subtle applications, like using a localized magnetic spot near a  nitrogen-carbon vacancy (NV-center)
to influence its magneto-optical  response, especially to increase its field-sensitivity at  certain applied field ranges.



\vspace{6pt} 




\acknowledgments{One of the authors (PDE) gratefully acknowledges discussions with M. Garc\'ia  (University of Kasel) and T. L\"uhmann (University of Leipzig). 
We thank N. Batova (Technological Institute) for the SEM measurements. This research was funded by the DFG under the grant DFG-ES 86/29-1 and DFG-ME 1564/11-1. The work in Russia was partially funded by RFBR and NSFC research project 20-52-53051. The research stay of M.T.G. was supported by the DAAD under the Research Stays for University Academics and Scientists 2021 programme, Nr. 57552334.}

\end{document}